# Optimized cerebral blood flow measurement in speckle contrast optical spectroscopy via refinement of noise calibration

**NINGHE LIU,**[†] **YU XI HUANG,**[†*] **SIMON MAHLER, CHANGHUEI YANG**[**]

*Department of Electrical Engineering, California Institute of Technology; Pasadena, CA 91125, USA.*
[†]*These authors contributed equally to this work.*
[*]*yuxi@caltech.edu*
[**]*chyang@caltech.edu*

**Speckle contrast optical spectroscopy (SCOS) offers a non-invasive and cost-effective method for monitoring cerebral blood flow (CBF). However, extracting accurate CBF from SCOS necessitates precise noise pre-calibration. Errors from this can degrade CBF measurement fidelity, particularly when the overall signal level is low. Such errors primarily stem from residual speckle contrast associated with camera and shot noise, whose fluctuations exhibit a temporal structure that mimics cerebral blood volume (CBV) waveforms. We propose an optimization-based framework that performs an adaptive refinement of noise calibration, mitigating the CBV-mimicking artifacts by reducing the CBF-CBV waveform correlation. Validated on 10 human subjects, our approach effectively lowered the signal threshold for reliable CBF signal from 97 to 26 electrons per pixel for a 1920x1200 pixels SCOS system. This improvement enables more accurate and robust CBF measurements in SCOS, especially at large source-detector (SD) distances for deeper tissue interrogation.**

Measurement of cerebral blood flow (CBF) is crucial for understanding brain function and pathology [1]. Speckle contrast optical spectroscopy (SCOS) [2–7], also known as speckle visibility spectroscopy (SVS) [8–11], has emerged as a promising non-invasive technique for this purpose, utilizing laser speckle contrast dynamics for its quantification [12]. This process yields a blood flow index (BFI) that is linearly correlated with the underlying blood flow [13,14]. In SCOS measurements, BFI is derived from the reciprocal of the measured spatial speckle contrast squared ($K^2$), following the proportionality $BFI \propto 1/K^2$ [13,14].

To achieve high brain specificity in CBF assessment, SCOS usually employs a large source-to-detector (S-D) distance [11,15], allowing light to predominantly probe deeper cerebral tissue rather than superficial scalp and skull layers [16–18], as illustrated in Fig. 1(a). By leveraging cost-effective CMOS sensors to simultaneously collect millions of speckles over a large detection area, SCOS achieves a high signal-to-noise ratio (SNR) [18,19]. This enhanced SNR makes SCOS particularly well-suited to overcome the signal loss at large S-D distances, thus enabling effective CBF measurements from deeper cerebral tissue [4,17,20–24].

Despite the inherent SNR advantages of SCOS's spatial sampling approach, the accuracy of its derived CBF critically depends on noise calibration [2,3]. Fundamentally, $K^2$ is defined as the ratio of the spatial variance of pixel intensities ($\sigma_I^2$) to the square of their mean intensity ($\langle I \rangle^2$):

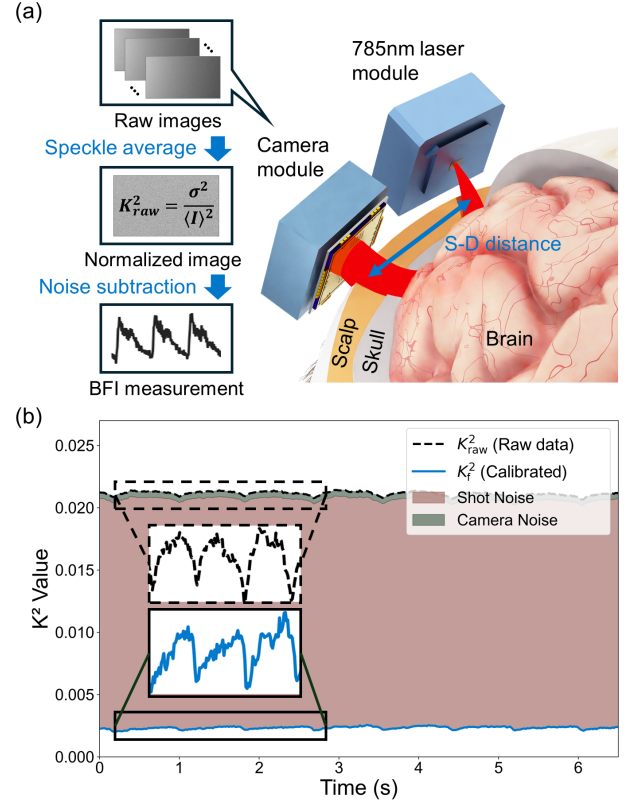

Fig. 1. (a) Compact SCOS device configuration and schematic of standard BFI processing pipeline. (b) Comparison of speckle contrast signal before ($K_{raw}^2$) and after ($K_f^2$) noise subtraction.

$$K^2 = \frac{\sigma_I^2}{\langle I \rangle^2} \quad (1)$$

However, the initially measured $K^2$ is a composite value, including not only the desired flow-induced signal but also contributions from multiple noise-related terms. Prominent among these are photon shot noise, speckle inhomogeneity, and various forms of detector noise such as read noise and dark current, all of which influence the $K^2$ calculated via Eq. (1). A common SCOS processing scheme, as illustrated in Fig. 1(a), first involves signal averaging to mitigate speckle inhomogeneity from non-uniform intensity profiles [2], yielding the normalized speckle contrast squared $K_{raw}^2$. Subsequently, speckle contrast contribution from photon shot noise and specific camera-related detector noises are addressed by subtraction of the corresponding terms to isolate the true flow-induced speckle contrast squared $K_f^2$. This crucial subtraction step requires precise, often context-specific, knowledge of parameters such as camera gain factors and dark noise characteristics. Obtaining and maintaining such precise parameter values, however, is a persistent challenge due to their sensitivity to dynamic operation conditions, including ambient temperature fluctuations, inherent inter-camera variability and camera gain drifts over time. As depicted in Fig. 1(b), speckle contrast associated with shot noise and camera noise from raw speckle images can contribute to a substantial portion of $K_{raw}^2$. Imperfect noise subtraction can therefore leave residual noise components comparable in magnitude to the blood flow-induced signal and compromise CBF fidelity. The impact of effective noise calibration is visually underscored by the distinct differences between the cardiac cycle waveforms observed pre- and post-calibration (Fig. 1(b), insets). This vulnerability to residual noise becomes particularly acute when targeting deeper cerebral tissue, a process that requires larger S-D separations and is thus inherently challenged by low signal levels.

To overcome these limitations, we introduce an optimization-based framework that performs a refinement of noise calibration and improves the robustness of CBF measurement in SCOS. A central insight underpinning our approach is the recognition that speckle contrast associated with residual noise from imperfect calibrations, notably camera and shot noise, leads to temporal fluctuations in the derived speckle contrast that mimic cerebral blood volume (CBV) signal. These artifacts increase the observed similarity between CBF and CBV waveforms, a characteristic we quantify using their high-pass correlation, termed the Volume-Flow Similarity Index (VFSI), as detailed in Algorithm 1. The algorithm adaptively refines the coefficients of different noise terms to decorrelate the CBF and CBV signals. In this Letter, we first validate our method across various signal regimes and show that it effectively suppresses the CBV-mimicking noise in the processed CBF waveforms. We then present results from 10 human subjects, showcasing marked improvements that enable reliable CBF measurements at larger S-D distances and lower signal levels. This advancement contributes to more accurate and robust CBF measurements in SCOS with computationally efficient post-processing, lowering barriers for broader adoption in clinical neuromonitoring.

In the noise subtraction step illustrated in Fig. 1(b), we apply the noise model for speckle contrast and isolate the desired flow-induced signal $K_f^2$ from $K_{raw}^2$ [2,3] as:

$$K_f^2 = K_{raw}^2 - K_{shot}^2 - K_{cam}^2. \quad (2)$$

Here $K_{shot}^2$ and $K_{cam}^2$ represent the shot noise and camera noise-related speckle contrast terms, respectively, which can be formulated as:

$$K_{shot}^2 = \frac{g}{\langle I \rangle}; K_{cam}^2 = \frac{\sigma_{cam}^2}{\langle I \rangle^2} \quad (3)$$

where $g$ denotes the camera gain and $\sigma_{cam}^2$ is the pre-characterized camera noise variance. Both $K_{shot}^2$ and $K_{cam}^2$ are critically dependent on the mean detected light intensity $\langle I \rangle$ [2,3], which itself fluctuates over time reflecting physiological dynamics such as the heart rate cardiac cycle. Similarly, CBV fluctuations are also modulated by $\langle I \rangle$ via the Beer-Lambert Law [25,26]:

$$\Delta CBV \propto \Delta OD = \ln\left(\frac{I_0}{\langle I \rangle}\right), \quad (4)$$

where $\Delta OD$ denotes the change in optical density due to blood absorption, and $I_0$ is the baseline intensity. This shared dependence on $\langle I \rangle$ is a key insight: if the pre-calibrated values for camera gain and noise variance contain residual errors $\Delta g$ and $\Delta \sigma_{cam}^2$ respectively, the inaccuracies in the subtracted noise-related terms, $\Delta K_{shot}^2$ and $\Delta K_{cam}^2$, are given by:

$$\Delta K_{shot}^2 = \frac{\Delta g}{\langle I \rangle}; \Delta K_{cam}^2 = \frac{\Delta \sigma_{cam}^2}{\langle I \rangle^2} \quad (5)$$

The intensity-dependence of these error terms, explicitly shown in Eq. (5), means that as $\langle I \rangle$ fluctuates with CBV, the errors also fluctuate in a similar manner. This systematically imprints a CBV-like temporal pattern onto the calculated $K_f^2$ and creates artifacts in the derived CBF. As $\langle I \rangle$ decreases towards lower signal levels, these structured noise-related terms become increasingly visible in the derived CBF waveform, as illustrated in Supplement 1, Note 1. To address such CBV-mimicking artifacts arising from imperfect noise parameter pre-calibration, we developed an optimization framework detailed in Algorithm 1. This algorithm adaptively refines the estimates for $g$ and $\sigma_{cam}^2$ by iteratively working to reduce the squared value of the Volume-Flow Similarity Index (VFSI[2]). Such optimization algorithm effectively decreases the artifactual similarity between the optimized CBF waveform and the concurrently measured CBV waveform, thus selectively mitigating the mis-calibrated noise components.

**Algorithm 1. CBF optimization framework**

**Input**: Time-sequence speckle contrast signal $K_{raw}^2$, average intensity $\langle I \rangle$, pre-calibrated camera gain $g_0$ and noise variance $\sigma_0^2$
**Output**: Calibrated CBF signal
**Initialization**: Initialize Adam optimizer and noise-related coefficients $g \leftarrow g_0$, $\sigma_{cam}^2 \leftarrow \sigma_0^2$
**Main algorithm**
Extract CBV waveform $CBV_{hp} = Highpass(CBV)$
For each iteration $i = 1, 2, \ldots, n$
    1. Adjust $K_f^2$ signal: $K_{af}^2 = K_{raw}^2 - g/\langle I \rangle - \sigma_{cam}^2/\langle I \rangle^2$
    2. Calculate CBF waveform: $CBF_{hp} = Highpass(1/K_{af}^2)$
    3. Calculate VFSI: $VFSI = corr(CBF_{hp}, CBV_{hp})$
    4. Update $g$ and $\sigma^2$ to reduce $loss = VFSI^2$
**Termination**
Return optimized $CBF^{opt} = 1/K_{af}^2$

To investigate SCOS performance across varying signal levels under controlled conditions, we acquired 20-second data traces from a subject's forehead using our previously developed SCOS device [2]. The detected signal level was adjusted by varying the laser source power from 100 mW to 65 mW in 5 mW increments. This setup allowed for precise modulation of the signal level (quantified as average electrons per pixel) by adjusting illumination laser power to eight distinct levels, while keeping the fixed 3 cm S-D distance and probing position. In the ideal situation with perfect noise calibration, decreasing signal levels should cause the CBF waveform to primarily exhibit increased random fluctuations without fundamentally altering its morphology, given the unchanged measurement geometry. However, as shown in Fig. 2(a), because standard pre-calibration typically leaves residual errors, the derived CBF waveforms increasingly mirror an inverted CBV waveform as the signal levels decrease, leading to larger VFSI$^2$. To quantify this CBF fidelity degradation, we correlated the CBF signal at each illumination level with that obtained at the highest signal level, as illustrated in Fig. 2(c). We also showed the VFSI$^2$ values across different signal levels in Fig. 2(d). For standard processing (blue dots in Fig. 2(c) and 2(d)), a sharp decline in fidelity and a rapid increase in VFSI$^2$ are observed below a signal threshold of approximately 90 electrons per pixel. In contrast, CBF waveforms derived using our proposed method (Fig. 2(b)) exhibit clear pulsatile traits, featuring accurate recovery of the systolic peak height and a clearly distinguishable dicrotic notch. Correspondingly, our method (red stars in Fig. 2(c) and 2(d)) maintains high CBF fidelity and low VFSI$^2$ across all tested signal levels. This indicates robust performance and a significantly reduced signal threshold for consistent CBF measurements in SCOS.

To further evaluate our optimization framework across a practical range of measurement conditions, we analyzed SCOS data from 10 human subjects recorded at varying S-D distances on the forehead. For each subject, three 20-second recordings were acquired at each S-D distance while keeping the laser illumination power constant. The S-D distance was increased from 3.0 cm until the signal level dropped below 10 electrons per pixel. As shown in Fig. 3(a), the detected signal strength diminishes with S-D distance following an exponential decay, which is fitted as a linear trend on a logarithmic signal axis. We then assessed data quality using VFSI$^2$ as an indicator of CBV-mimicking artifact levels relative to signal strength. For data processed with standard methods (Fig. 3(b)), a piece-wise linear fit applied to the VFSI$^2$ versus signal strength data indicates a signal threshold of approximately 97 electrons per pixel; below this threshold, VFSI$^2$ values rapidly increase, signifying severely compromised data quality. This observed signal threshold and the trend of escalating VFSI$^2$ at lower signal strengths are consistent with the performance of standard SCOS processing previously shown in Fig. 2. In contrast, applying our optimization framework (Fig. 3(d)) yields a significantly reduced signal threshold of approximately 26 electrons per pixel, demonstrating robust performance at significantly lower signal levels. Figure 3(c) presents illustrative examples of this improvement for two representative subjects, which correspond to points 1 and 2 highlighted in Fig. 3(b) and 3(d). These measurements were conducted at a 4.5 cm S-D distance, resulting in challenging low signal levels of 46 and 27 electrons per pixel, respectively. Before optimization, the derived CBF waveforms are heavily distorted and exhibit high absolute VFSI values (-0.61 and -0.69). After optimization, the CBF waveforms display clear physiological features, such as distinct systolic peaks and dicrotic notches, with substantially reduced absolute VFSI values (-0.17 and -0.01), visually confirming the enhanced CBF signal fidelity. These results demonstrate that our proposed method significantly expands the operational range of SCOS, enabling more reliable and consistent CBF measurements at larger S-D distances critical for deep brain tissue detection.

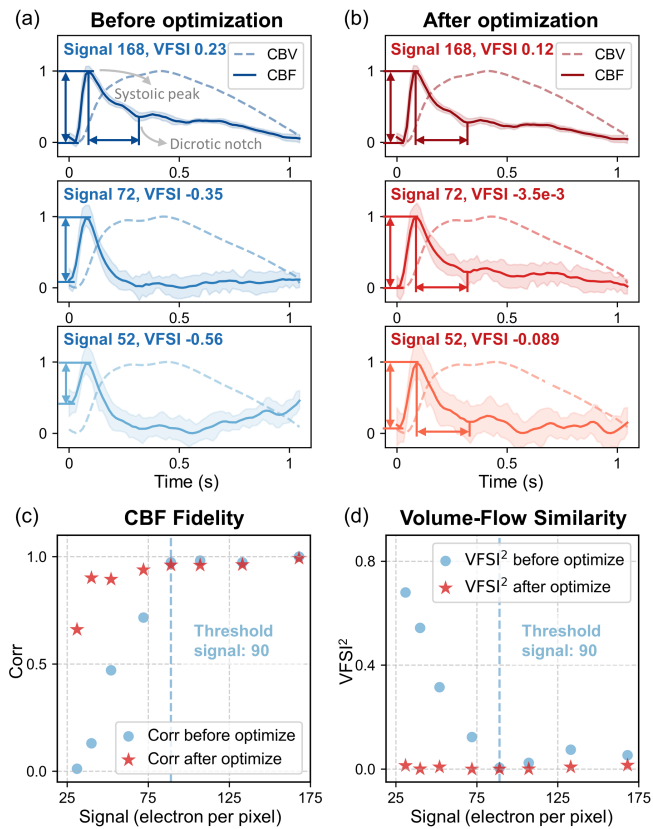

Fig. 2. Comparison of CBF measurements in SCOS using standard pre-calibration and the proposed optimization framework at varying signal strengths. Representative CBF and CBV waveforms are shown for (a) standard processing and (b) our optimization scheme; central lines denote mean waveforms over the 20-second recordings, with shaded areas representing ±1 standard deviation. Plots show (c) CBF fidelity and (d) Volume-Flow Similarity Index across varying signal strengths.

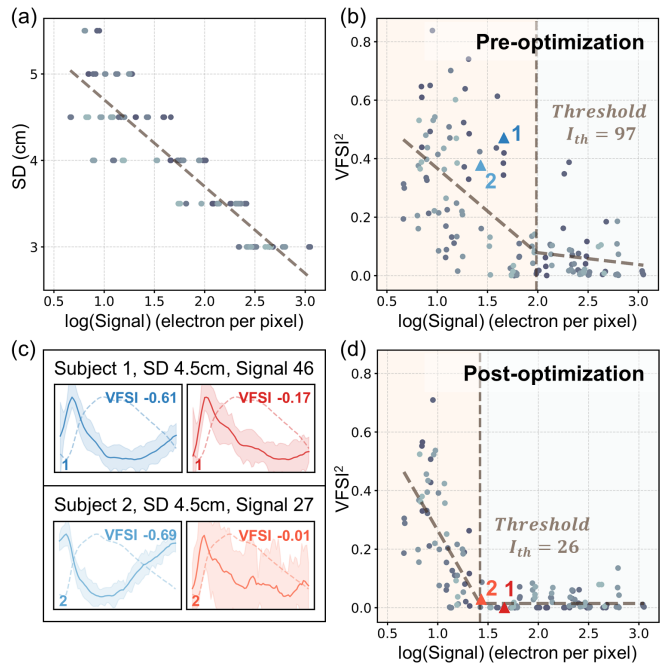

Fig. 3. Enhanced SCOS performance using the proposed algorithm across varying S-D separations and signal strengths in 10 human subjects. (a) Detected signal strength versus S-D separation. (b, d) VFSI$^2$ versus signal strength pre- (b) and post-optimization (d). The signal threshold determined by piece-wise linear fitting is reduced from 97 to 26 electrons per pixel. (c) Representative CBF and CBV waveforms for two highlighted subjects at a 4.5 cm S-D separation before and after optimization.

We also showcase the robustness of the proposed optimization framework by applying it to cases where a priori estimation of noise parameters is entirely absent, see Supplement 1, Note 2. The adaptive nature of the algorithm allows it to effectively auto-calibrate camera gain and noise variance by directly constraining the similarity between the derived CBF and concurrently measured CBV waveforms. As demonstrated with SCOS data acquired at a challenging low signal strength (4 cm S-D separation, 55 electrons per pixel), our method successfully retrieved reliable CBF signals even when the initial pre-calibrated values $g_0$ and $\sigma_0^2$ were set to zero. Beyond its robustness to unknown priors, the framework is computationally efficient: processing a 20-second SCOS time trace through 500 iterations of our Algorithm 1 took approximately 0.5 seconds on a commercial GPU (NVIDIA GTX 1080 Ti).

In conclusion, this Letter presents and validates an optimization-based framework for the adaptive refinement of noise calibration in SCOS. This approach effectively mitigates CBV-mimicking artifacts that stem from noise mis-calibration, enhancing the fidelity and robustness of derived CBF signals at large S-D distances (i.e. low signal levels). We have demonstrated its capability to extend the reliable operational range of SCOS, particularly for challenging low-light conditions critical for deeper brain detection, and to perform robustly even with minimal or absent a priori knowledge of underlying noise parameters. This computationally efficient method significantly reduces dependence on meticulous and often difficult-to-maintain noise pre-calibrations, thereby broadening its applicability across diverse SCOS instrumentations in clinical and potentially commercial applications.

**References**

1. J. A. H. R. Claassen, D. H. J. Thijssen, R. B. Panerai, and F. M. Faraci, Physiological Reviews **101**, 1487 (2021).
2. Y. X. Huang, S. Mahler, M. Dickson, A. Abedi, J. M. Tyszka, Y. T. Lo, J. Russin, C. Liu, and C. Yang, JBO **29**, 067001 (2024).
3. B. Kim, S. Zilpelwar, E. J. Sie, F. Marsili, B. Zimmermann, D. A. Boas, and X. Cheng, Commun Biol **6**, 1 (2023).
4. S. Mahler, Y. X. Huang, M. Ismagilov, D. Álvarez-Chou, A. Abedi, J. M. Tyszka, Y. T. Lo, J. Russin, R. L. Pantera, C. Liu, and C. Yang, NPh **12**, 015003 (2025).
5. C. P. Valdes, H. M. Varma, A. K. Kristoffersen, T. Dragojevic, J. P. Culver, and T. Durduran, Biomed. Opt. Express, BOE **5**, 2769 (2014).
6. C.-H. P. Lin, I. Orukari, C. Tracy, L. K. Frisk, M. Verma, S. Chetia, T. Durduran, J. W. Trobaugh, and J. P. Culver, Opt. Lett. **48**, 1427 (2023).
7. C. G. Favilla, S. Carter, B. Hartl, R. Gitlevich, M. T. Mullen, A. G. Yodh, W. B. Baker, and S. Konecky, Neurophoton. **11**, (2024).
8. R. Bandyopadhyay, A. S. Gittings, S. S. Suh, P. K. Dixon, and D. J. Durian, Review of Scientific Instruments **76**, 093110 (2005).
9. Y. X. Huang, S. Mahler, J. Mertz, and C. Yang, Opt. Express **31**, 31253 (2023).
10. J. Xu, A. K. Jahromi, J. Brake, J. E. Robinson, and C. Yang, APL Photonics **5**, 126102 (2020).
11. S. Mahler, Y. X. Huang, M. Liang, A. Avalos, J. M. Tyszka, J. Mertz, and C. Yang, Biomed. Opt. Express **14**, 4964 (2023).
12. D. A. Boas and A. K. Dunn, J. Biomed. Opt. **15**, 011109 (2010).
13. C. Liu, K. Kılıç, S. E. Erdener, D. A. Boas, and D. D. Postnov, Biomed. Opt. Express **12**, 3571 (2021).
14. D. D. Postnov, J. Tang, S. E. Erdener, K. Kılıç, and D. A. Boas, Sci. Adv. **6**, eabc4628 (2020).
15. Y. X. Huang, S. Mahler, M. Dickson, A. Abedi, Y. T. Lo, P. D. Lyden, J. Russin, C. Liu, C. Yang, and 10.48550/ARXIV.2501.19005, (2025).
16. G. E. Strangman, Z. Li, and Q. Zhang, PLoS ONE **8**, e66319 (2013).
17. W. Zhou, O. Kholiqov, J. Zhu, M. Zhao, L. L. Zimmermann, R. M. Martin, B. G. Lyeth, and V. J. Srinivasan, Sci. Adv. **7**, eabe0150 (2021).
18. M. B. Robinson, T. Y. Cheng, M. Renna, M. M. Wu, B. Kim, X. Cheng, D. A. Boas, M. A. Franceschini, and S. A. Carp, Neurophoton. **11**, (2024).
19. J. Xu, A. K. Jahromi, and C. Yang, APL Photonics **6**, 016105 (2021).
20. S. A. Carp, M. B. Robinson, and M. A. Franceschini, Neurophoton. **10**, (2023).
21. K. C. Wu, A. Martin, M. Renna, M. Robinson, N. Ozana, S. A. Carp, and M. A. Franceschini, Neurophoton. **10**, (2023).
22. C. G. Favilla, G. L. Baird, K. Grama, S. Konecky, S. Carter, W. Smith, R. Gitlevich, A. Lebron-Cruz, A. G. Yodh, and R. A. McTaggart, J NeuroIntervent Surg jnis (2024).
23. B. (Kenny) Kim, A. C. Howard, T. Y. Cheng, J. E. Anderson, B. B. Zimmermann, E. Hazen, L. Carlton, M. Robinson, M. Renna, M. A. Yucel, S. A. Carp, M. A. Franceschini, D. A. Boas, and X. Cheng, (2025).
24. Y. X. Huang, S. Mahler, A. Abedi, J. M. Tyszka, Y. T. Lo, P. D. Lyden, J. Russin, C. Liu, and C. Yang, Biomedical Optics Express **15**, 6083 (2024).
25. W. B. Baker, A. B. Parthasarathy, D. R. Busch, R. C. Mesquita, J. H. Greenberg, and A. G. Yodh, Biomed Opt Express **5**, 4053 (2014).
26. I. Oshina and J. Spigulis, J Biomed Opt **26**, 100901 (2021).

**Full References**

1. J. A. H. R. Claassen, D. H. J. Thijssen, R. B. Panerai, and F. M. Faraci, "Regulation of cerebral blood flow in humans: physiology and clinical implications of autoregulation," Physiological Reviews **101**, 1487–1559 (2021).
2. Y. X. Huang, S. Mahler, M. Dickson, A. Abedi, J. M. Tyszka, Y. T. Lo, J. Russin, C. Liu, and C. Yang, "Compact and cost-effective laser-powered speckle contrast optical spectroscopy fiber-free device for measuring cerebral blood flow," JBO **29**, 067001 (2024).
3. B. Kim, S. Zilpelwar, E. J. Sie, F. Marsili, B. Zimmermann, D. A. Boas, and X. Cheng, "Measuring human cerebral blood flow and brain function with fiber-based speckle contrast optical spectroscopy system," Commun Biol **6**, 1–10 (2023).
4. S. Mahler, Y. X. Huang, M. Ismagilov, D. Álvarez-Chou, A. Abedi, J. M. Tyszka, Y. T. Lo, J. Russin, R. L. Pantera, C. Liu, and C. Yang, "Portable six-channel laser speckle system for simultaneous measurement of cerebral blood flow and volume with potential applications in characterization of brain injury," NPh **12**, 015003 (2025).
5. C. P. Valdes, H. M. Varma, A. K. Kristoffersen, T. Dragojevic, J. P. Culver, and T. Durduran, "Speckle contrast optical spectroscopy, a non-invasive, diffuse optical method for measuring microvascular blood flow in tissue," Biomed. Opt. Express, BOE **5**, 2769–2784 (2014).
6. C.-H. P. Lin, I. Orukari, C. Tracy, L. K. Frisk, M. Verma, S. Chetia, T. Durduran, J. W. Trobaugh, and J. P. Culver, "Multi-mode fiber-based speckle contrast optical spectroscopy: analysis of speckle statistics," Opt. Lett. **48**, 1427 (2023).
7. C. G. Favilla, S. Carter, B. Hartl, R. Gitlevich, M. T. Mullen, A. G. Yodh, W. B. Baker, and S. Konecky, "Validation of the Openwater wearable optical system: cerebral hemodynamic monitoring during a breath-hold maneuver," Neurophoton. **11**, (2024).
8. R. Bandyopadhyay, A. S. Gittings, S. S. Suh, P. K. Dixon, and D. J. Durian, "Speckle-visibility spectroscopy: A tool to study time-varying dynamics," Review of Scientific Instruments **76**, 093110 (2005).
9. Y. X. Huang, S. Mahler, J. Mertz, and C. Yang, "Interferometric speckle visibility spectroscopy (iSVS) for measuring decorrelation time and dynamics of moving samples with enhanced signal-to-noise ratio and relaxed reference requirements," Opt. Express **31**, 31253 (2023).
10. J. Xu, A. K. Jahromi, J. Brake, J. E. Robinson, and C. Yang, "Interferometric speckle visibility spectroscopy (ISVS) for human cerebral blood flow monitoring," APL Photonics **5**, 126102 (2020).
11. S. Mahler, Y. X. Huang, M. Liang, A. Avalos, J. M. Tyszka, J. Mertz, and C. Yang, "Assessing depth sensitivity in laser interferometry speckle visibility spectroscopy (iSVS) through source-to-detector distance variation and cerebral blood flow monitoring in humans and rabbits," Biomed. Opt. Express **14**, 4964 (2023).
12. D. A. Boas and A. K. Dunn, "Laser speckle contrast imaging in biomedical optics," J. Biomed. Opt. **15**, 011109 (2010).
13. C. Liu, K. Kılıç, S. E. Erdener, D. A. Boas, and D. D. Postnov, "Choosing a model for laser speckle contrast imaging," Biomed. Opt. Express **12**, 3571 (2021).
14. D. D. Postnov, J. Tang, S. E. Erdener, K. Kılıç, and D. A. Boas, "Dynamic light scattering imaging," Sci. Adv. **6**, eabc4628 (2020).
15. Y. X. Huang, S. Mahler, M. Dickson, A. Abedi, Y. T. Lo, P. D. Lyden, J. Russin, C. Liu, C. Yang, and 10.48550/ARXIV.2501.19005, "Assessing Sensitivity of Brain-to-Scalp Blood Flows in Laser Speckle Imaging by Occluding the Superficial Temporal Artery," (2025).
16. G. E. Strangman, Z. Li, and Q. Zhang, "Depth Sensitivity and Source-Detector Separations for Near Infrared Spectroscopy Based on the Colin27 Brain Template," PLoS ONE **8**, e66319 (2013).
17. W. Zhou, O. Kholiqov, J. Zhu, M. Zhao, L. L. Zimmermann, R. M. Martin, B. G. Lyeth, and V. J. Srinivasan, "Functional interferometric diffusing wave spectroscopy of the human brain," Sci. Adv. **7**, eabe0150 (2021).
18. M. B. Robinson, T. Y. Cheng, M. Renna, M. M. Wu, B. Kim, X. Cheng, D. A. Boas, M. A. Franceschini, and S. A. Carp, "Comparing the performance potential of speckle contrast optical spectroscopy and diffuse correlation spectroscopy for cerebral blood flow monitoring using Monte Carlo simulations in realistic head geometries," Neurophoton. **11**, (2024).
19. J. Xu, A. K. Jahromi, and C. Yang, "Diffusing wave spectroscopy: A unified treatment on temporal sampling and speckle ensemble methods," APL Photonics **6**, 016105 (2021).
20. S. A. Carp, M. B. Robinson, and M. A. Franceschini, "Diffuse correlation spectroscopy: current status and future outlook," Neurophoton. **10**, (2023).
21. K. C. Wu, A. Martin, M. Renna, M. Robinson, N. Ozana, S. A. Carp, and M. A. Franceschini, "Enhancing diffuse correlation spectroscopy pulsatile cerebral blood flow signal with near-infrared spectroscopy photoplethysmography," Neurophoton. **10**, (2023).
22. C. G. Favilla, G. L. Baird, K. Grama, S. Konecky, S. Carter, W. Smith, R. Gitlevich, A. Lebron-Cruz, A. G. Yodh, and R. A. McTaggart, "Portable cerebral blood flow monitor to detect large vessel occlusion in patients with suspected stroke," J NeuroIntervent Surg jnis-2024-021536 (2024).
23. B. (Kenny) Kim, A. C. Howard, T. Y. Cheng, J. E. Anderson, B. B. Zimmermann, E. Hazen, L. Carlton, M. Robinson, M. Renna, M. A. Yucel, S. A. Carp, M. A. Franceschini, D. A. Boas, and X. Cheng, "Mapping human cerebral blood flow with high-density, multi-channel speckle contrast optical spectroscopy," (2025).
24. Y. X. Huang, S. Mahler, A. Abedi, J. M. Tyszka, Y. T. Lo, P. D. Lyden, J. Russin, C. Liu, and C. Yang, "Correlating stroke risk with non-invasive cerebrovascular perfusion dynamics using a portable speckle contrast optical spectroscopy laser device," Biomedical Optics Express **15**, 6083–6097 (2024).
25. W. B. Baker, A. B. Parthasarathy, D. R. Busch, R. C. Mesquita, J. H. Greenberg, and A. G. Yodh, "Modified Beer-Lambert law for blood flow," Biomed Opt Express **5**, 4053–4075 (2014).
26. I. Oshina and J. Spigulis, "Beer–Lambert law for optical tissue diagnostics: current state of the art and the main limitations," J Biomed Opt **26**, 100901 (2021).